\documentclass[preprintnumbers,
amsmath,amssymb,floatfix,10pt,prd,onecolumn,
superscriptaddress,longbibliography,nofootinbib]{revtex4-2}
\usepackage{bm}
\usepackage{amsfonts}
\usepackage{latexsym}
\usepackage{amsmath}
\usepackage{textcomp}
\usepackage{float}
\usepackage{booktabs}
\usepackage{dcolumn}
\usepackage{dsfont}
\usepackage{ragged2e}
\usepackage{epsfig}
\usepackage[dvipsnames]{xcolor}
\usepackage{hyperref}
\hypersetup{           
	colorlinks=true,                
	breaklinks=true,                
	urlcolor= OrangeRed,                
	linkcolor= magenta,                
	bookmarksopen=false,
	filecolor=black,
	citecolor=red,
	linkbordercolor=blue}
\usepackage{graphicx}
\usepackage{overpic}
\usepackage{orcidlink}
\usepackage[T1]{fontenc}

\begin{document}
	\title{Exploring Rényi Entropic Cosmology Constrained by DESI DR2 BAO and Complementary Late-Time Observations}

   \author{Rajdeep Mazumdar \orcidlink{0009-0003-7732-875X}}
	\email[Corresponding author: ]{rajdeepmazumdar377@gmail.com}
	\affiliation{%
		Department of Physics, Dibrugarh University, Dibrugarh, Assam, India, 786004}

  \author{Kalyan Malakar\orcidlink{0009-0002-5134-1553}}%
\email{kalyanmalakar349@gmail.com}
\affiliation{Department of Physics, Dibrugarh University, Dibrugarh, Assam, India, 786004}
\affiliation{Department of Physics, Silapathar College, Dhemaji, Assam, India, 787059}
	
	\author{Kalyan Bhuyan\orcidlink{0000-0002-8896-7691}}%
	\email{kalyanbhuyan@dibru.ac.in}
	\affiliation{%
		Department of Physics, Dibrugarh University, Dibrugarh, Assam, India, 786004}%
	\affiliation{Theoretical Physics Divison, Centre for Atmospheric Studies, Dibrugarh University, Dibrugarh, Assam, India 786004}
	
\begin{abstract}
We investigate the late-time cosmological viability of Rényi entropic cosmology (REC) by confronting its modified Friedmann dynamics with DESI DR2 BAO, cosmic chronometers, gravitational-wave standard sirens, redshift-space distortions, and two independent Type Ia supernova compilations, PantheonPlus and Union3. Unlike earlier works, we obtain a stringent constraint on the Rényi parameter $\lambda \sim (3.7\text{--}3.8)\times10^{-124}$, with the resulting value even satisfying recent Big Bang Nucleosynthesis and baryogenesis bounds. Across all observational combinations, REC yields $H_0\simeq68.2$--$68.4$ $km\,s^{-1}\,Mpc^{-1}$ and $\Omega_{m0}\simeq0.291$--$0.295$, with a transition from deceleration to acceleration at $z_{\rm tr}\simeq0.66$--$0.68$. The effective equation of state remains quintessence-like, with $w_{\rm eff}(0)\simeq-0.68$, and does not cross the phantom boundary. REC provides a modest improvement in the minimum $\chi^2$ relative to $\Lambda$CDM, while the AIC, DIC, BIC, and Bayesian evidence indicate that the preference for REC is statistically competitive but not decisive. Conclusively, the results show that Rényi entropy can produce a controlled modification of the late-time expansion history while remaining observationally close to $\Lambda$CDM.
\end{abstract}
	\maketitle
	
\section{Introduction}	
The Universe is currently undergoing an accelerated phase of expansion, as demonstrated by measurements from Type Ia supernovae (SN Ia)~\cite{riess1998,perlmutter1999}, the Cosmic Microwave Background (CMB)~\cite{bennett2003,spergel2003}, Baryon Acoustic Oscillations (BAO)~\cite{eisenstein2005,percival2010}, and Large-Scale Structure (LSS) surveys, among other recent developments in observational cosmology. Dark energy (DE), an enigmatic component that makes up around 70\% of the universe's total energy density, is frequently attributed as a cause for this late-time acceleration. Despite its remarkable success in explaining observational data, the fundamental nature and origin of dark energy remain elusive from a theoretical perspective. The cosmological constant ($\Lambda$), which serves as the foundation for the conventional $\Lambda$CDM model, is the most straightforward and commonly recognised candidate for dark energy. Excellent agreement with a wide variety of cosmological observations from various eras has been shown by this hypothesis. Nevertheless, the $\Lambda$CDM framework has a number of serious conceptual problems even with its practical effectiveness. The most well-known of these is the cosmological constant issue, which results from the huge difference between the observed and theoretically expected levels of vacuum energy density\cite{weinberg1989,Martin2012}. Further challenging our comprehension are the fine-tuning and cosmic coincidence problems~\cite{weinberg1989,carroll1992,sahni2000}, which ask why the current value of dark energy density is so perfectly tuned and why its dominance corresponds with the current period of cosmic evolution. Such theoretical challenges show the need for exploration of other theoretical possibilities causing the late-time cosmic acceleration, outside the domain of standard $\Lambda$CDM paradigm. Broadly, such approaches can be classified into two main categories. The first involves introducing dynamical dark energy components with evolving equations of state, including models such as quintessence, phantom fields, k-essence~\cite{ArmendarizPicon2000}, tachyon fields~\cite{Sen2002}, chameleon fields~\cite{Khoury2004}, and Chaplygin gas models~\cite{Kamenshchik2001,Bento2002}. The second approach focuses on modifying the underlying theory of gravity itself, leading to various extended or alternative gravitational frameworks that aim to naturally explain the observed acceleration without invoking an explicit dark energy component.\\
In recent years due to insights from black hole physics, an alternative and conceptually intriguing approach has emerged from the interplay between gravity and thermodynamics. Since Bardeen \textit{et al.} developed black hole mechanics and introduced Bekenstein–Hawking entropy~\cite{bardeen1973,bekenstein1973,hawking1975}, the relationship between spacetime dynamics and thermodynamic principles has been more apparent. Concepts like the holographic principle, gauge/gravity duality, and quantum information-theoretic approaches to gravity have further reinforced this connection~\cite{tHooft1993,susskind1995,maldacena1999,almheiri2013}.
Jacobson's groundbreaking study showed that the first rule of thermodynamics applied to local Rindler horizons can yield Einstein's field equations~\cite{jacobson1995}, indicating that gravity itself might be an emergent phenomenon.  Further, the Friedmann equations can be found by applying thermodynamic rules to the apparent horizon of a Friedmann–Lemaître–Robertson–Walker (FLRW) universe~\cite{cai2005,akbar2006}. Such a thermodynamic approach has been effectively extended to a broad class of modified gravity theories~\cite{sheykhi2018,nojiri2017} and offers a coherent framework in which cosmic dynamics are guided by horizon thermodynamics. Modifications to the entropy–area relation naturally result in different cosmic dynamics under this thermodynamic framework. A number of generalised entropy formalisms, including Tsallis, Rényi, Barrow, Kaniadakis, and Sharma–Mittal entropies, as well as corrections inspired by loop quantum gravity~\cite{tsallis1988,renyi1961,barrow2020,kaniadakis2002,sharma1975,rovelli1996}, have been proposed in response to the long-range nature of gravity and potential quantum gravitational effects. These generalised entropies capture non-extensive and information-theoretic aspects of gravitational systems and, in suitable situations, reduce to the standard Bekenstein–Hawking entropy. They result in modified Friedmann equations and effective energy components that can drive acceleration in both early and late times when applied to the thermodynamics of cosmic horizons. These modifications typically take the shape of logarithmic and power-law changes. Specifically, the effects of long-range correlations are captured by power-law corrections that result from the entanglement of quantum fields both inside and outside the horizon \cite{x13,x14}. However, the uncertainty principle, quantum fluctuations, and thermal equilibrium fluctuations in loop quantum gravity naturally lead to logarithmic corrections \cite{x16,x21}. In quantum information and communication theory, the Shannon entropy, which is a member of the class of logarithmic entropies, is essential and meets a number of crucial operational requirements \cite{x22}. However, Shannon entropy might not be sufficient to describe non-asymptotic regimes, where the rule of big numbers is not applicable. Alternative entropy metrics, including collision entropy and other generalised non-extensive formulations, become more significant in these situations. Motivated by these considerations, Rényi~\cite{renyi1961} introduced a generalized entropy formalism that unifies various entropy measures within a single consistent framework, where the entropy associated with the horizon can be expressed as \cite{renyi1961}:
\begin{equation}
S_R = \frac{1}{\lambda} \ln\left(1 + \lambda S_{BH}\right),
\end{equation}
\label{E1}
where, $S_{BH}$ is the standard Bekenstein–Hawking entropy, and $\lambda$ is the Rényi parameter that characterises deviations from the standard entropy. The conventional thermodynamic description is recovered when the Rényi entropy drops to the standard Bekenstein–Hawking entropy in the limiting scenario $\lambda \to 0$.\\
Rényi entropy \cite{renyi1961}, originally introduced by Alfréd Rényi, serves as a one-parameter generalization of the Boltzmann–Gibbs entropy and provides a consistent framework to incorporate non-extensive and information-theoretic effects into gravitational dynamics. A number of studies have explored its implications across different cosmic epochs. For instance, investigations of Rényi entropy in Big Bang Nucleosynthesis (BBN) and baryogenesis \cite{n1} have placed important constraints on the Rényi parameter, demonstrating its viability in describing early Universe thermodynamics without conflicting with primordial element abundances. Furthermore, works on Rényi entropy-corrected expanding universe models \cite{n2} have shown that such corrections can significantly affect the evolution of the Hubble parameter and deceleration parameter, offering distinguishable signatures from standard $\Lambda$CDM cosmology. In addition, Rényi entropy has been explored in the framework of inflationary cosmology \cite{n3}, where it provides alternative mechanisms for realizing inflationary dynamics consistent with observational constraints. The connection between Rényi entropy and holographic dark energy models \cite{n4} has also been widely studied, particularly in flat spacetime scenarios, where it modifies the infrared cutoff and leads to improved agreement with observational data. More recently, Rényi entropy has been proposed as a tool to quantify cosmic homogeneity \cite{n5}, linking information-theoretic measures with large-scale structure formation. These diverse applications collectively highlight that Rényi entropy not only offers a unifying thermodynamic description but also serves as a powerful phenomenological framework to probe deviations from standard cosmology across both early- and late-time regimes.\\
Despite these promising developments, a comprehensive assessment of Rényi entropic cosmological (REC) model, together with direct and stringent constraints on the Rényi parameter $\lambda$ in the light of latest high-precision observational data remains relatively unexplored. Recent advances in observational cosmology have significantly strengthened constraints on theoretical models of cosmic acceleration. The high-precision BAO measurements from DESI DR2~\cite{ref90} provide stringent tests of extensions beyond $\Lambda$CDM, including dynamical dark energy and modified gravity~\cite{refk58,refk65,refk68,refk69}. Gravitational-wave standard sirens~\cite{g1,g2} provide an independent probe of the expansion history through luminosity-distance measurements, while cosmic chronometers (CC)~\cite{PK34,PK35,PK36,PK37} and redshift-space distortions (RSD)~\cite{PK41} constrain the expansion rate and growth of structure, respectively. Independent Type~Ia supernova compilations, such as PantheonPlus~\cite{PK45} and Union3~\cite{PK48}, further provide complementary constraints and enable assessment of the dependence on the adopted SN dataset. Motivated by this, it is both timely and essential to investigate the viability of REC model at late times and to place stringent observational constraints on its parameters using these recent high-precision datasets. Such an analysis can provide deeper insights into the role of entropy-based modifications in cosmic evolution and assess their potential as viable alternatives to the standard $\Lambda$CDM framework.\\
The paper is structured as: Section \ref{s2} explains Rényi entropic cosmology, Section \ref{sec3} explains the observational data and methodology employed to constraint the model, Section \ref{s4} discuses the constraints and cosmological implications obtained, Section \ref{s5} gives a summary of our results.
\section{Rényi Entropic Cosmology}\label{s2}
We consider a spatially homogeneous and isotropic Universe described by the Friedmann–Lemaître–Robertson–Walker (FLRW) metric \cite{renyi1961,n1},
\begin{equation}
ds^2 = h_{\mu\nu} dx^\mu dx^\nu + R^2 \left(d\theta^2 + \sin^2\theta\, d\phi^2 \right).
\end{equation}
Here, $R = a(t) r$, $x^0 = t$, and $x^1 = r$, with the two-dimensional metric taking the form $h_{\mu\nu} = \mathrm{diag}\left(-1, \frac{a^2}{1 - k r^2}\right)$. The spatial curvature parameter $k = -1, 0, +1$ represents open, flat, and closed geometries, respectively. In this framework, we choose to apply thermodynamics to the apparent horizon. While alternative horizons exist (such as the Hubble or future event horizons), the apparent horizon is conventionally employed as a working ansatz in standard entropic cosmology, though it does not represent a unique thermodynamic boundary. The radius of the apparent horizon can be expressed as\cite{renyi1961,n1,n2}:
\begin{equation}
R = \frac{1}{\sqrt{H^2 + \frac{k}{a^2}}},
\end{equation}
where $H = \dot{a}/a$ denotes the Hubble parameter. The corresponding temperature, given in terms of the surface gravity, is expressed as:
\begin{equation}
T_h = \frac{\kappa}{2\pi} = -\frac{1}{2\pi R} \left(1 - \frac{\dot{R}}{2HR}\right).
\end{equation}
In the quasi-static limit $\dot{R} \ll 2HR$, one recovers the standard Bekenstein-Hawking temperature $T_h = 1/(2\pi R)$. We acknowledge that modifying the entropy function generally requires a re-derivation of the energy-temperature relation; however, for the purposes of this phenomenological model, we retain the standard temperature definition as our primary assumption. Taking the cosmic fluid to be a perfect fluid with the energy–momentum tensor defined as,
\begin{equation}
T_{\mu\nu} = (\rho + p) u_\mu u_\nu + p g_{\mu\nu},
\end{equation}
where the energy density and pressure of the cosmic fluid are $\rho$ and $p$, respectively. Conservation of energy–momentum\cite{renyi1961,n1},
$\nabla_\mu T^{\mu\nu} = 0$, leads to the continuity equation:
\begin{equation}
\dot{\rho} + 3H(\rho + p) = 0.
\label{E2}
\end{equation}
The work density is defined using the standard projection of the stress-energy tensor:
\begin{equation}
W = -\frac{1}{2} T^{\mu\nu} h_{\mu\nu} = \frac{1}{2} (\rho - p).
\end{equation}
Applying the first law of thermodynamics on the apparent horizon \cite{renyi1961,n1,n2},
\begin{equation}
dE = T_h dS_h + W dV,
\end{equation}
where $E = \frac{4\pi}{3} \rho R^3$ and $V = \frac{4\pi}{3} R^3$, we obtain:
\begin{equation}
dE = 4\pi R^2 \rho\, dR - 4\pi H R^3 (\rho + p)\, dt.
\end{equation}
To incorporate non-extensive effects, we consider the Rényi entropy (Eq. \ref{E1}) associated with the horizon. Assuming deviations from the standard entropy are small ($\lvert \lambda S \rvert \ll 1$), we expand to leading order \cite{renyi1961,n1}:
\begin{equation}
S_h = S - \frac{\lambda}{2} S^2,
\end{equation}
where $S = \frac{A}{4G} = \frac{\pi R^2}{G}$. Its differential form is
\begin{equation}
dS_h = (1 - \lambda S)\, dS,
\end{equation}
with
\begin{equation}
dS = \frac{2\pi R \dot{R}}{G} dt.
\end{equation}
Substituting the above relations into the first law and simplifying, we obtain \cite{renyi1961,n1}:
\begin{equation}
(1 - \lambda S)\frac{dR}{R^3} = 4\pi G H (\rho + p)\, dt.
\end{equation}
Using the continuity equation, this can be rewritten as \cite{renyi1961}:
\begin{equation}
-\frac{2\, dR}{R^3} \left(1 - \alpha R^2 \right) = \frac{8\pi G}{3} d\rho,
\end{equation}
where $\alpha = \frac{\lambda \pi}{G}$. Integrating, we obtain
\begin{equation}
\frac{1}{R^2} - \alpha \ln\left(\frac{C}{R^2}\right) = \frac{8\pi G}{3} \rho,
\end{equation}
where $C$ is an integration constant having the same dimensions as $R^2$. In this modified framework, $C$ introduces a new phenomenological degree of freedom. To ensure a non-pathological expansion history, we require $CH^2 > 0$. Substituting $R^{-2} = H^2 + \frac{k}{a^2}$, the modified Friedmann equation becomes:
\begin{equation}
H^2 + \frac{k}{a^2} - \alpha \ln\left(C\left(H^2 + \frac{k}{a^2}\right)\right) = \frac{8\pi G}{3} \rho.
\end{equation}
For $\alpha \to 0$, the above equation reduces to the standard form \cite{n1}. For a flat Universe $(k=0)$, this reduces to \cite{n1}:
\begin{equation}
H^2 - \alpha \ln \left(C H^2 \right) = \frac{8\pi G}{3} \rho.
\label{E3}
\end{equation}
Using the standard definitions of the matter and radiation density parameters, the modified Friedmann equation given by Eq.~(\ref{E3}) can be recast in
physical units as
\begin{equation}
\left(\frac{H}{H_0}\right)^{2} = \Omega_{m0} (1+z)^{3} + \Omega_{r0} (1+z)^{4} + \frac{ {\pi} \lambda \ln \left(C H^2\right)}{G {H_0}^{2}},
\label{E4x}
\end{equation}
where $H_0$ is the present-day Hubble parameter, while $\Omega_{m0}$ and $\Omega_{r0}$ denote the present-day matter and radiation density parameters, respectively. The argument of the logarithm, $C H^2$, is dimensionless, with $C$ having dimensions of $H^{-2}$. Employing the boundary condition $H(z=0)=H_0$ in Eq. (\ref{E4x}), we can express the parameter $\lambda$ as:
\begin{equation}
\lambda = \frac{G H_0^{2}\left(1-\Omega_{m0}-\Omega_{r0}\right)}{\pi \ln\!\left(C H_0^2\right)}.
\label{E5x}
\end{equation}
Upon adopting the natural-unit convention $8\pi G=1$, Eq.~(\ref{E4x}) reduces to:
\begin{equation}
\left(\frac{H}{H_0}\right)^{2} = \Omega_{m0} (1+z)^{3} + \Omega_{r0} (1+z)^{4} + \frac{ 8{\pi}^{2} \lambda \ln \left(C H^2\right)}{ {H_0}^{2}},
\label{E4}
\end{equation}
And, Eq. (\ref{E4x}) reduces to:
\begin{equation}
\lambda = \frac{H_0^{2}\left(1-\Omega_{m0}-\Omega_{r0}\right)}{8\pi^2 \ln\!\left(C H_0^2\right)}.
\label{E5}
\end{equation}
Substituting Eq. (\ref{E5}) back into Eq. (\ref{E4}), we obtain the final background evolution equation:
\begin{equation}
\left(\frac{H(z)}{H_0}\right)^{2} = \Omega_{m0} (1 + z)^3 + \Omega_{r0} (1 + z)^4 + \left(1-\Omega _{m0}-\Omega _{r0}\right) \frac{\ln \left(C H^2\right)}{\ln \left(C H_0^2\right)}.
\label{E6}
\end{equation}
Equation~(\ref{E6}) is subsequently used to determine the background expansion history and the corresponding cosmological observables, which are then confronted with the observational datasets described in the following sections.

\section{Methodology and Observational Data}
\label{sec3}
The free parameters of the Rényi entropic cosmological (REC) model are constrained within a Bayesian framework using the Markov Chain Monte Carlo (MCMC) method. We employ the affine-invariant ensemble sampler implemented in the \texttt{emcee} package to efficiently sample the posterior distributions, assuming uniform priors over physically motivated parameter ranges. At each MCMC step, the cosmological evolution is solved numerically to evaluate the theoretical predictions corresponding to the observational data. After discarding the initial burn-in phase, the converged chains are analyzed using \texttt{GetDist} to obtain the marginalized posterior distributions together with the corresponding confidence intervals. Specifically, we employ the latest DESI DR2 baryon acoustic oscillation (BAO) measurements, observational Hubble parameter measurements obtained from the cosmic chronometer (CC) technique, and gravitational-wave (GW) standard sirens. In addition, Type Ia supernova observations from the PantheonPlus and Union3 compilations are analyzed separately to assess the robustness of the inferred cosmological constraints. The combination of these independent probes enables a comprehensive test of the model over a broad redshift range while significantly reducing parameter degeneracies. The observational datasets and the corresponding statistical analysis are as given below:
\begin{enumerate}
\item{\textbf{Cosmic Chronometers}: Cosmic Chronometers provide direct measurements of the Hubble expansion rate by exploiting the differential age evolution of passively evolving galaxies. Since this technique is largely independent of any fiducial cosmological model, it offers one of the most reliable probes of the expansion history. We employ the latest compilation of 34 CC measurements spanning the redshift range $0.07<z<1.965$  from the compilations in Refs.~\cite{PK34,PK35,PK36,PK37},which provides robust constraints on the background evolution of the Universe. Ref.~\cite{PK38} provides a comprehensive description for the formulation of the $\chi^{2}$ function for this data, while Ref.~\cite{PK39} provides the full covariance 
matrix associated with the correlated CC data.}
\item{\textbf{DESI DR2 BAO}: To probe the expansion history over intermediate and high redshifts, we employ the latest baryon acoustic oscillation (BAO) measurements from the DESI Data Release 2 (DR2) \cite{ref90}. The dataset comprises observations of more than 14 million galaxies and quasars distributed across seven effective redshift bins spanning $0.295 \leq z \leq 2.334$. It includes four tracer samples, namely the Bright Galaxy Sample (BGS), Luminous Red Galaxies (LRGs), Emission Line Galaxies (ELGs), and Quasars (QSOs) \cite{AK1,PKx}. The BGS provides an isotropic BAO measurement at an effective redshift of $z=0.295$, while the LRG, ELG, and QSO samples yield anisotropic BAO measurements over the range $0.51<z<2.33$. These observations constrain the volume-averaged distance, transverse comoving distance, and Hubble distance relative to the sound horizon, thereby providing stringent constraints on the late-time expansion history \cite{AK1,PKx}. Owing to its unprecedented statistical precision and broad redshift coverage, DESI DR2 offers one of the most powerful probes for testing dark energy and modified cosmological models. Refs~\cite{AK1,PKx} provides detailed explanation about the construction of the $\chi^{2}$ function and the full covariance matrix associated with this data.}
\item{\textbf{Redshift-Space Distortions}: The Redshift-Space Distortion (RSD) measurements, probes the growth of large-scale structures through the quantity $f\sigma_8(z)$ \cite{PK41}. The RSD signal originates from the peculiar velocities of galaxies, which introduce anisotropies in the observed galaxy clustering pattern in redshift space. Consequently, RSD measurements are sensitive to both the cosmic expansion history and the growth rate of matter perturbations, making them an important probe for distinguishing modified cosmological models from the standard $\Lambda$CDM scenario \cite{PK41}. In this work, we employ the latest compilation of RSD measurements obtained from several galaxy surveys, including WiggleZ, BOSS, eBOSS, SDSS, and VIPERS, covering a wide range of redshifts as given in Ref.~\cite{PK40}. Since these observations provide information complementary to geometrical probes, their inclusion helps reduce parameter degeneracies and leads to tighter constraints on the REC model.}
\item{\textbf{PantheonPlus Type Ia Supernovae}: To constrain the late-time expansion history, we employ the PantheonPlus compilation of Type Ia supernovae, comprising 1701 spectroscopically confirmed events over the redshift range $0.001<z<2.26$ \cite{PK45}. The dataset combines observations from multiple surveys and is uniformly reanalyzed within the SALT2 framework, incorporating improved photometric calibration, light-curve standardization, selection effects, and systematic uncertainties \cite{PK46}. Consequently, PantheonPlus provides one of the most precise and homogeneous supernova samples currently available for cosmological studies. Following Refs.~\cite{PKx,PKx1,DJ1}, we use the uncalibrated PantheonPlus sample by excluding the SH0ES Cepheid-calibrated supernovae and analytically marginalizing over the supernova absolute magnitude. This treatment avoids imposing an external prior on the Hubble constant, thereby ensuring consistency with DESI DR2 BAO measurements, which constrain only relative distance scales and are insensitive to the absolute distance scale \cite{PK47}. This well lines with the baseline approach adopted in DESI DR2 cosmological analyses \cite{ref90}. As a result, the combined analysis yields unbiased constraints on the late-time cosmological evolution and the parameters of the REC model.}
\item{\textbf{Union3 Type Ia Supernovae}: To examine the robustness of the cosmological constraints, we also consider the recently released Union3 compilation of Type Ia supernovae \cite{PK48}. The dataset consists of 2087 supernovae spanning the redshift range $0.001<z<2.26$, making it one of the largest currently available supernova samples. Union3 employs an independent Bayesian hierarchical framework for light-curve standardization and the treatment of systematic uncertainties, thereby providing a complementary dataset to PantheonPlus \cite{PK48}. Owing to its independent calibration strategy and improved control of systematics, Union3 serves as an important cross-check for assessing the stability of the inferred cosmological parameters within the REC model. Ref.~ \cite{PK49} presents a comprehensive methodological discussion associated to this data.}
\item{\textbf{Gravitational-Wave Standard Sirens}: We further incorporate gravitational-wave (GW) observations as an independent and complementary probe of the cosmic expansion history through standard sirens. Specifically, we use publicly available luminosity distance measurements from the LIGO--Virgo--KAGRA observing runs O1, O2, and O3, as reported in the GWTC-1, GWTC-2, GWTC-2.1, and GWTC-3 catalogues~\cite{g1,g2}. The analysis employs the median luminosity distances together with their corresponding $1\sigma$ uncertainties obtained from Bayesian parameter estimation. The source redshifts are determined either from identified electromagnetic counterparts or from statistically inferred host-galaxy associations for binary black hole mergers. Since gravitational-wave standard sirens provide absolute luminosity distance measurements without relying on the traditional cosmic distance ladder, they offer an independent test of the late-time expansion history. When combined with DESI DR2 BAO, cosmic chronometers, redshift-space distortion measurements, and Type Ia supernova observations, GW data help break parameter degeneracies and significantly improve the constraints on the REC model \cite{PKx1,DJ1}.}
\end{enumerate}
The model parameters are constrained by maximizing the joint likelihood constructed from the observational datasets considered in this work. Since these datasets are assumed to be statistically independent, the total likelihood $\mathcal{L}_{\rm tot}(\mathbf{\theta})$ is given by the product of individual likelihoods $\mathcal{L}_i(\mathbf{\theta})$, corresponding to a total chi-square ($\chi^2$) defined by:
\begin{equation}
    \chi^2_{\rm tot}(\mathbf{\theta}) = -2 \ln \mathcal{L}_{\rm tot}(\mathbf{\theta}) = \sum_{i} \chi^2_i(\mathbf{\theta}).
\end{equation}
Depending on the dataset combination under consideration, the total chi-square is expressed as:
\begin{equation}
\chi^2_{\rm tot} =
\begin{cases}
\chi^2_{\rm DESI}
+\chi^2_{\rm CC}
+\chi^2_{\rm GW}
+\chi^2_{\rm RSD},
& \text{DESI + CC + GW + RSD}, \\[8pt]

\chi^2_{\rm DESI}
+\chi^2_{\rm CC}
+\chi^2_{\rm GW}
+\chi^2_{\rm RSD}
+\chi^2_{\rm PantheonPlus},
& \text{DESI + CC + GW + RSD + PantheonPlus}, \\[8pt]

\chi^2_{\rm DESI}
+\chi^2_{\rm CC}
+\chi^2_{\rm GW}
+\chi^2_{\rm RSD}
+\chi^2_{\rm Union3},
& \text{DESI + CC + GW + RSD + Union3}.
\end{cases}
\label{eq:chi2_total}
\end{equation}
Throughout this work, we assume a spatially flat Universe ($k=0$) and fix the present radiation density parameter to $\Omega_{r0}=9.2\times10^{-5}$. Since the DESI DR2 BAO measurements constrain cosmological distances relative to the comoving sound horizon at the drag epoch, the sound horizon scale $r_d$ is treated as a free parameter and marginalized over during the parameter estimation rather than being fixed to a fiducial value. Likewise, because redshift-space distortion (RSD) observations probe the growth of cosmic structures, the parameter $\sigma_{8,0}$ which denotes the present-day root-mean-square amplitude of linear matter density fluctuations on a scale of $8\,h^{-1}\,\mathrm{Mpc}$ is also treated as a free-parameter for the model. Consequently, for the REC model, we constrain the 5-dimensional parameter space $\mathbf{\theta} = \{H_0, \Omega_{m0}, C, r_d, \sigma_{8,0}\}$ using flat priors: $H_0 \in [60, 80]$, $\Omega_{m0} \in [0, 1]$, $C \in [0, 2]$, $r_d \in [130, 165]$, and $\sigma_{8,0} \in [0.5, 1.1]$. The MCMC chains are initialized at $\mathbf{\theta}_0 = (70.0, 0.3, 1.0, 147.0, 0.81)$. To quantify the goodness of fit and model complexity, we employ the Akaike Information Criterion (AIC), Bayesian Information Criterion (BIC), and Deviance Information Criterion (DIC), defined as
\begin{equation}
\mathrm{AIC}=\chi^2_{\min}+2k,\qquad
\mathrm{BIC}=\chi^2_{\min}+k\ln N,
\end{equation}
\begin{equation}
\mathrm{DIC}=2\bar D-D(\bar{\boldsymbol{\theta}}),
\qquad
D(\boldsymbol{\theta})=-2\ln\mathcal L(\boldsymbol{\theta}),
\end{equation}
where $k$ is the number of free parameters ($k=5$ for the REC model and $k=4$ for $\Lambda\mathrm{CDM}$), $N$ is the total number of data points, $\bar D=\langle D\rangle$ is the posterior-averaged deviance, and $D(\bar{\boldsymbol{\theta}})$ is evaluated at the posterior-mean parameters. Here, $\chi^2_{\min}$ is obtained from the maximum-likelihood solution. We further compute the Bayesian
evidence $\mathcal{Z}$, based on which the relative preference between the REC and $\Lambda$CDM model is quantified through
the logarithmic Bayes factor, defined as $\ln B=\ln \mathcal{Z}_{\Lambda\mathrm{CDM}}-\ln \mathcal{Z}_{\mathrm{REC}}$, such that $\ln B>0$ ($\ln B<0$) favours $\Lambda\mathrm{CDM}$ (REC). For the information criteria, we adopt
\begin{equation}
\Delta X=X_{\Lambda\mathrm{CDM}}-X_{\mathrm{REC}},
\qquad
X\in{\chi^2_{\min},\mathrm{AIC},\mathrm{BIC},\mathrm{DIC}},
\end{equation}
with $\Delta X>0$ in case of $\chi^2_{\min}$ indicates the REC model provides a better fit to the data than the $\Lambda$CDM model, and $\Delta X>0$ in case of $\mathrm{AIC},\mathrm{BIC},\mathrm{DIC}$ indicates a preference for the REC model after accounting for its additional free parameter, whereas $\Delta X <0$ favor the $\Lambda$CDM model. Here, AIC emphasizes predictive fit with a relatively modest complexity penalty, BIC imposes a stronger sample size dependent penalty for additional parameters, while DIC accounts for the effective complexity of the model through its posterior distribution.

\section{Constraints and Cosmological Implications}\label{s4}
\begin{figure*}[htb]
\centerline{
\includegraphics[width=1.05\textwidth]{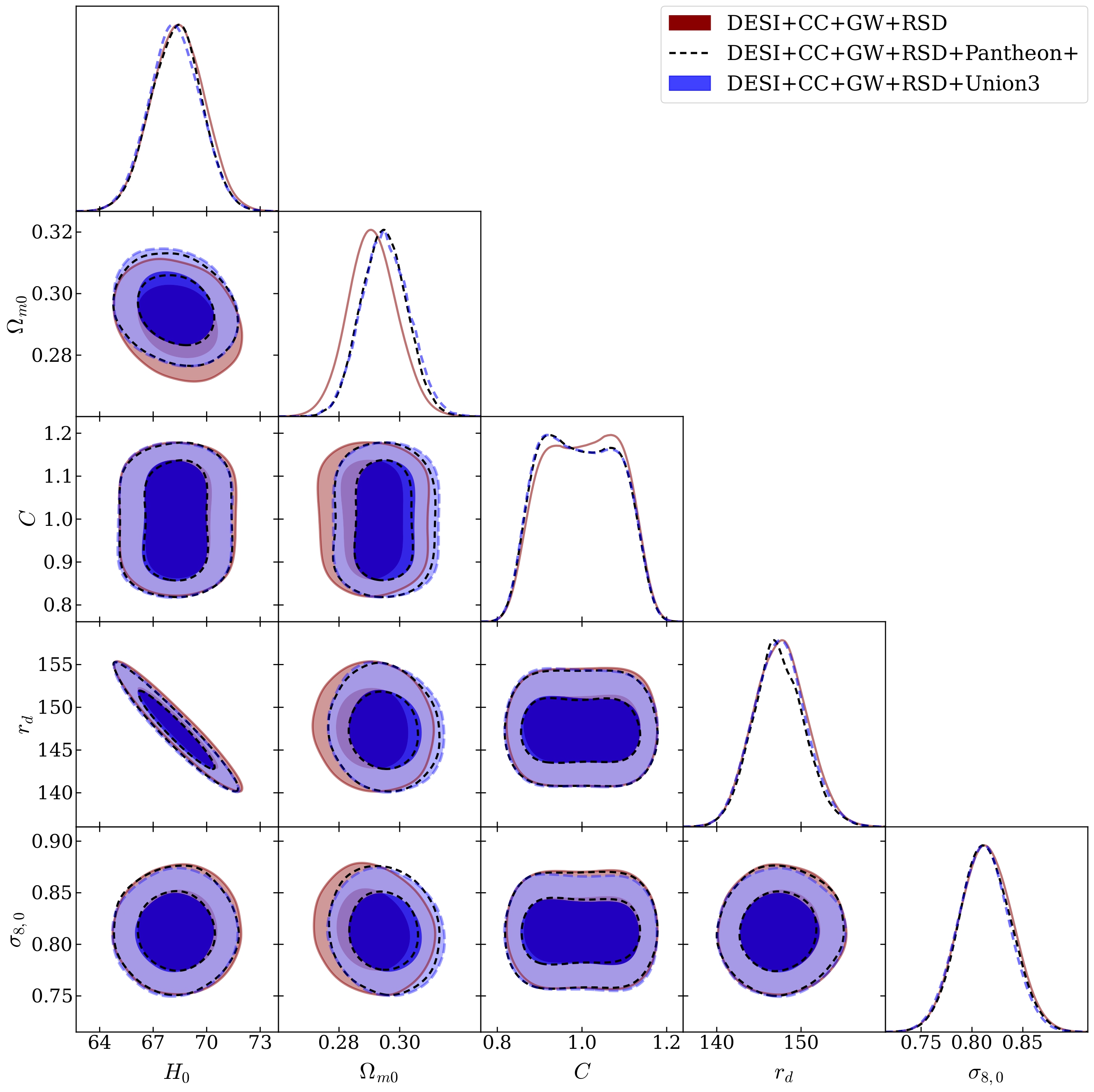}}
\caption{ 2D likelihood contours for the parameter space of REC model, showing the 1-$\sigma$ and 2-$\sigma$ confidence regions (corresponding to 68\% and 95\% confidence levels).}
\label{f1}
\end{figure*}

\begin{figure*}[htb]
\centerline{
\includegraphics[width=1.05\textwidth]{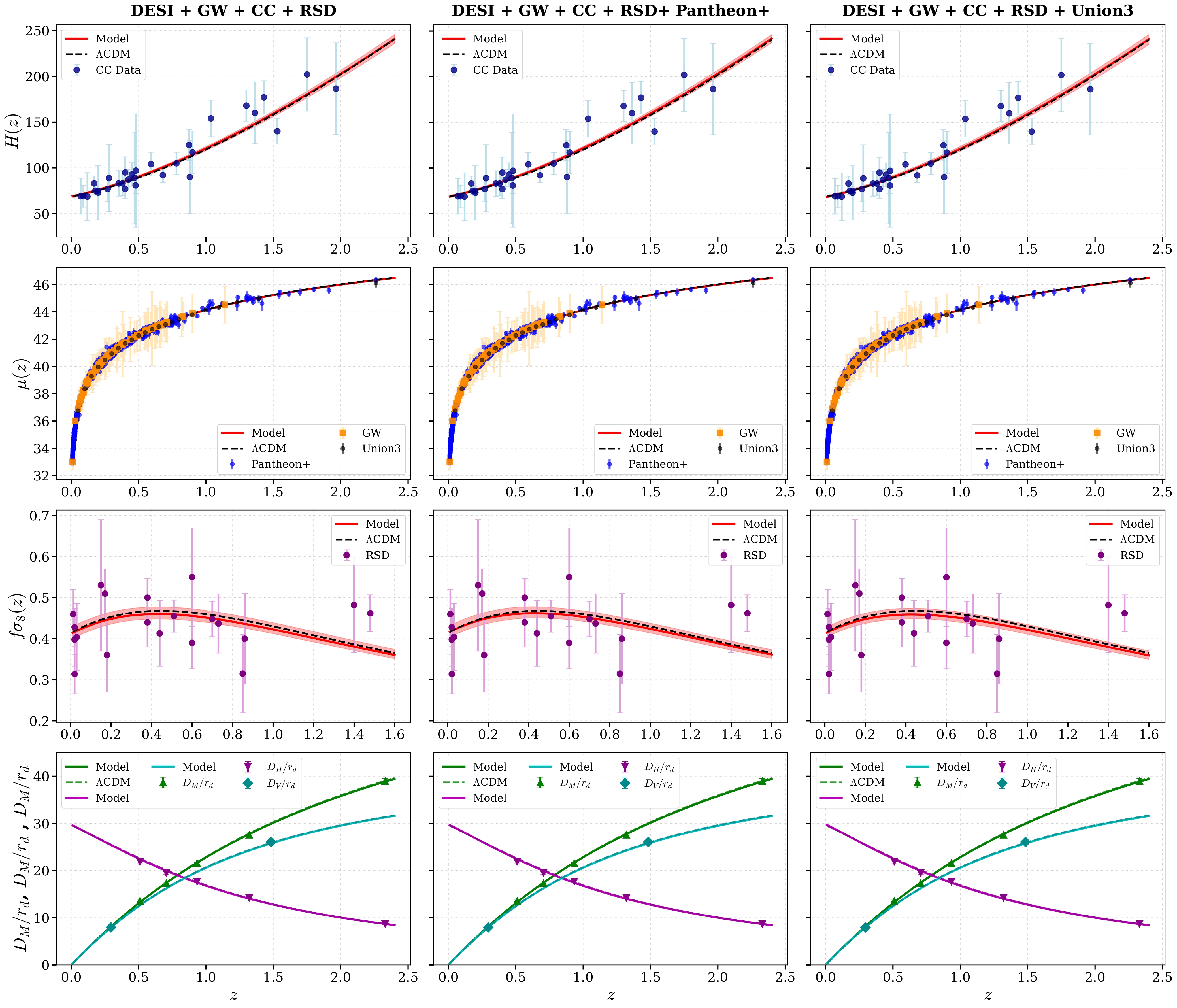}}
\caption{Redshift evolution of the model predictions for the best-fit parameter values, shown together with the corresponding observational data. For comparison, the predictions of the standard $\Lambda$CDM model are also included in each panel.}
\label{f2}
\end{figure*}

\begin{figure*}[htb]
\centerline{
\includegraphics[width=1\textwidth]{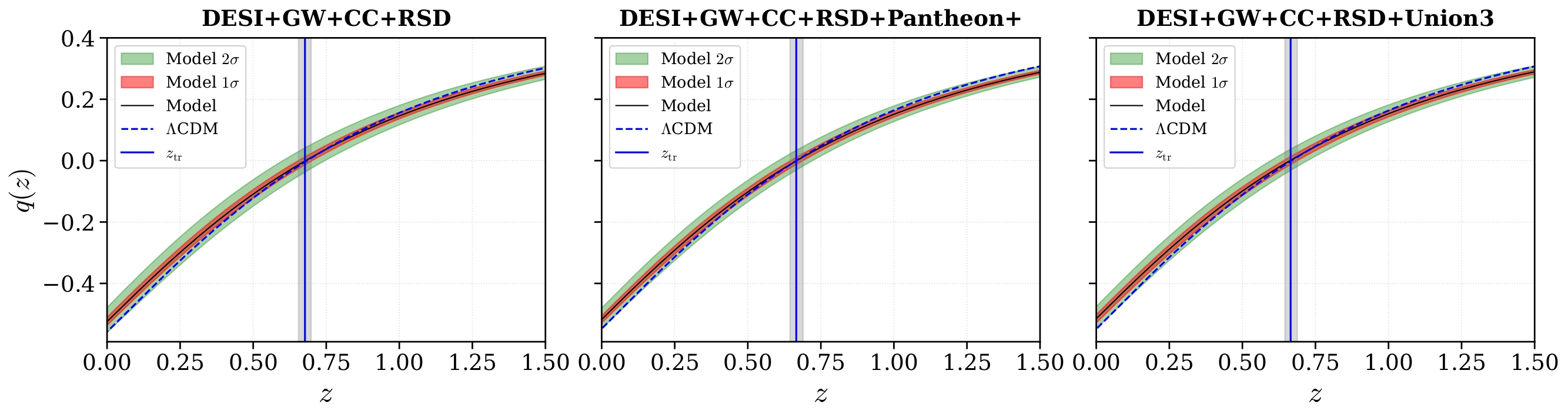}}
\centerline{
\includegraphics[width=1\textwidth]{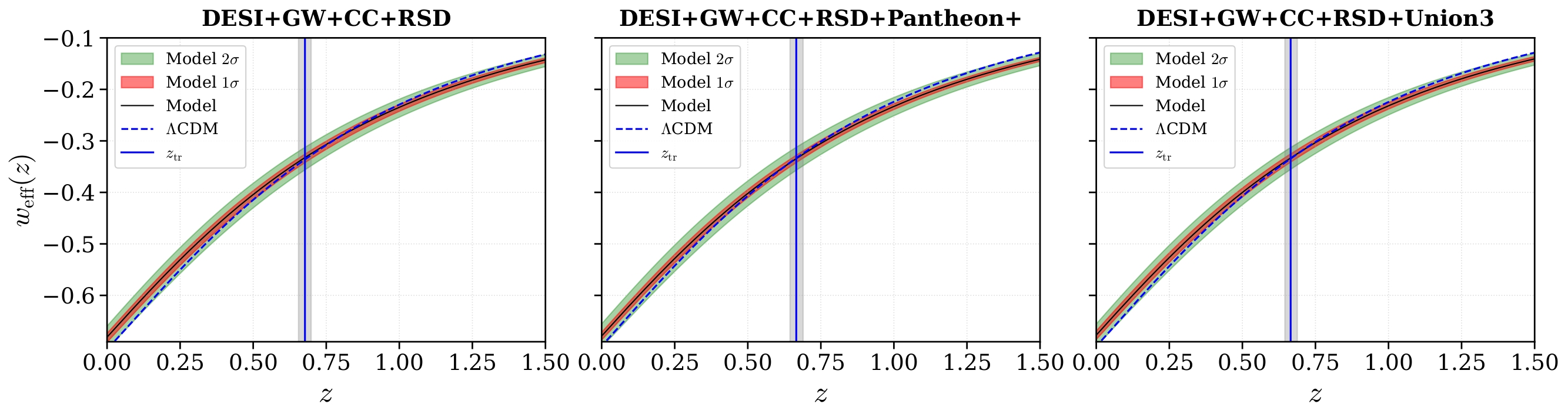}}
\centerline{
\includegraphics[width=1\textwidth]{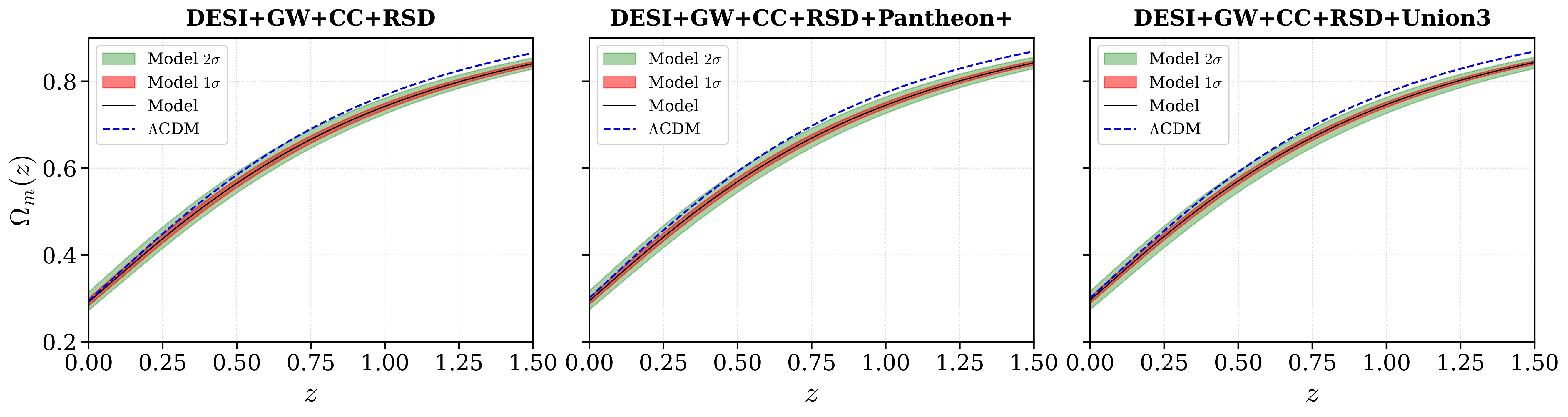}}

\centerline{
\includegraphics[width=1\textwidth]{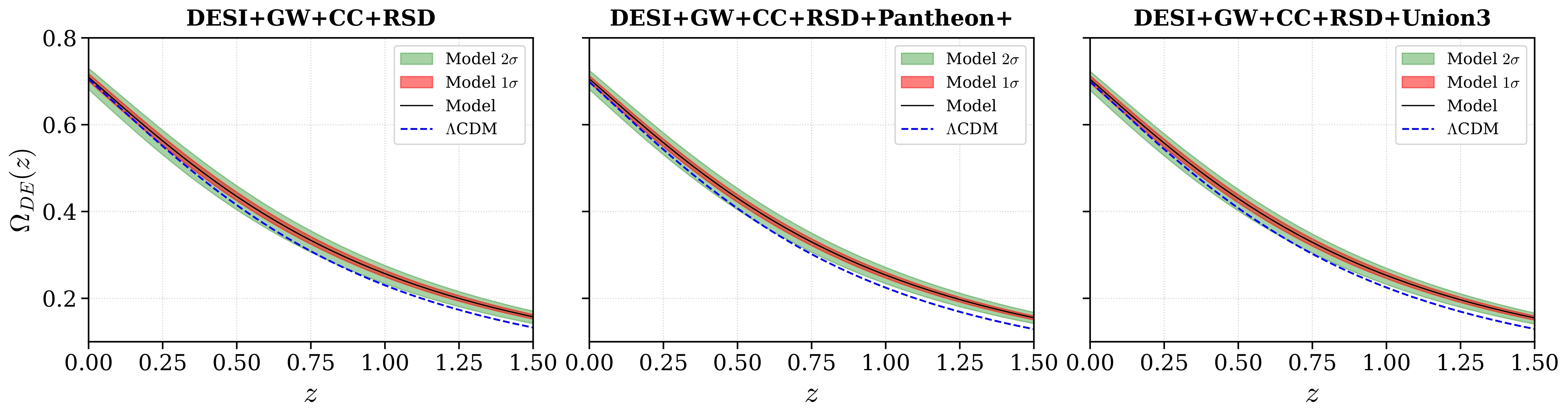}}

\caption{Plot of the evolution of deceleration parameter, density parameter, and effective EoS parameter for the best fit values of the REC model.}
\label{f3}
\end{figure*}

\begin{figure*}[htb]
\centerline{
\includegraphics[width=0.9\textwidth]{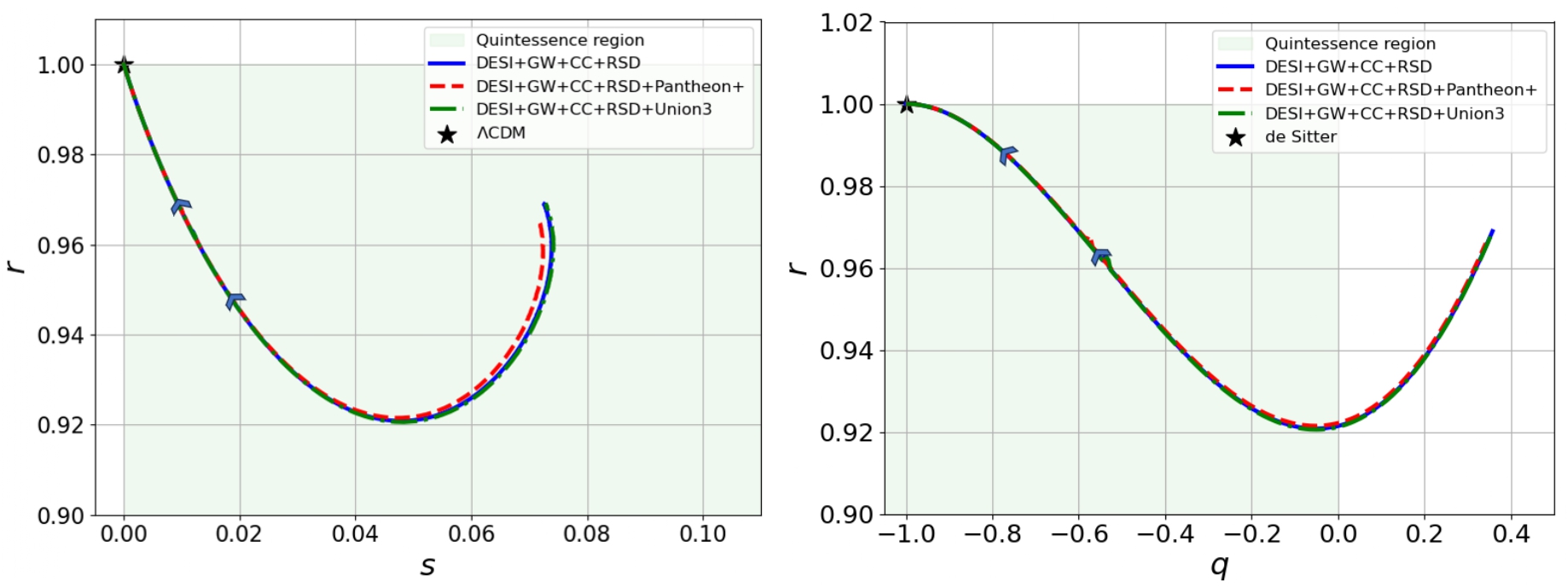}}
\caption{Evolution of the statefinder diagnostic pairs $(r,s)$ and $(r,q)$ with redshift for the best-fit parameter values obtained from different datasets. The arrowheads indicate the direction of evolution of the trajectories.}
\label{f4}
\end{figure*}

Based on the observational datasets and statistical methodology described in the previous sections, we now present the parameter constraints and cosmological implications of the proposed REC model. We consider three combinations of late-time cosmological observations, namely DESI+GW+CC+RSD, DESI+GW+CC+RSD+PantheonPlus, and DESI+GW+CC+RSD+Union3. The resulting constraints on the model parameters are summarized in Table~\ref{tab1}, while a statistical comparison with the standard $\Lambda$CDM model is provided in Table~\ref{tab2}. The corresponding constraints on the present-day values of the cosmological parameters are reported in Table~\ref{tab3}. The marginalized posterior distributions of the model parameters, together with the redshift-dependent evolution of the model predictions against the corresponding observational data, are shown in Figs.~(\ref{f1}) and~ (\ref{f2}), respectively. We further examine the resulting cosmic dynamics through the redshift evolution of key kinematic and dynamical quantities, including the deceleration parameter, effective equation-of-state (EoS) parameter, and density parameter as shown in Fig. (\ref{f3}).\\
\begin{enumerate}
\item{\textbf{Parameter Estimation and Cosmological Tensions:} The results shows that the inferred parameters remain stable across all three dataset combinations. Combining the results from DESI+GW+CC+RSD with PantheonPlus and Union3, the REC model gives $H_0\simeq68.24$--$68.42$ $km\,s^{-1}\,Mpc^{-1}$ and $\Omega_{m0}\simeq0.291$--$0.295$, compared with $H_0\simeq68.65$--$68.92$ $km\,s^{-1}\,Mpc^{-1}$ and $\Omega_{m0}\simeq0.294$--$0.301$ for $\Lambda$CDM. Thus, REC model consistently prefers a slightly lower $H_0$ and matter density than $\Lambda$CDM, although the differences remain well within the inferred uncertainties. Interestingly, in REC model the additional model parameter is remarkably stable, with $C\simeq0.99$--$1.00$ across all dataset combinations, indicating that the preferred model realization is largely insensitive to the choice of supernova compilation. The sound-horizon constraint is likewise highly stable, with $r_d\simeq147.21$--$147.36$~$Mpc$ for REC model compared with $147.37$--$147.57$~$Mpc$ for $\Lambda$CDM. This near invariance indicates that the modification primarily affects the late-time expansion history rather than producing a substantial shift in the drag-epoch scale. In contrast, the clustering amplitude is systematically higher in REC model, with $\sigma_{8,0}\simeq0.813$--$0.815$, compared with $\sigma_{8,0} \simeq0.794$--$0.796$ for $\Lambda$CDM. Therefore, although the modified model improves the overall fit to the combined late-time observations, it does not suppress the amplitude of matter clustering and hence does not alleviate the $S_{8}\equiv\sigma_{8}\sqrt{\Omega_{m}/0.3}$ tension within the present observational framework \cite{S11,S12}. The inferred $H_0$ indicates that the REC model does not significantly alleviate the $H_0$ tension~\cite{DJ1,verde2019tensions}. Its preferred range, $H_0\simeq68.2$--$68.4$~$km\,s^{-1}\,Mpc^{-1}$, remains close to the $\Lambda$CDM value, lies well below the local distance-ladder measurement, $H_0=73.04\pm1.04$~$km\,s^{-1}\,Mpc^{-1}$~\cite{S13}, and is slightly above the Planck constraint, $H_0=67.4\pm0.5$~$km\,s^{-1}\,Mpc^{-1}$~\cite{S14}. Hence, REC mildly shifts the inferred expansion rate but does not resolve the $H_0$ tension.}
\item{\textbf{Statistical Comparison with $\Lambda$CDM:} To assess whether the additional degree of freedom introduced by the REC model is statistically supported by the observations, we compare its performance with the standard $\Lambda$CDM model using $\chi^2_{\min}$, AIC, BIC, DIC, and the Bayesian evidence. The corresponding results are summarized in Table~\ref{tab2}. Across all three dataset combinations, REC model consistently yields a lower minimum $\chi^2$, with an improvement of $\Delta\chi^2\simeq0.86$--$3.52$ relative to $\Lambda$CDM. The improvement is modest for DESI+GW+CC+RSD alone but becomes more pronounced after including either PantheonPlus or Union3, indicating that the supernova data enhance the relative goodness of fit of the modified model. Subsequently, the difference in AIC for  DESI+GW+CC+RSD is negative, it lies in the range $\Delta{\rm AIC}\simeq 1.48$--$1.52$ after the inclusion of PantheonPlus or Union3, while $\Delta{\rm BIC}\simeq-4.18$ to $-1.66$ throughout. Thus, the AIC result evolves from a slight preference for $\Lambda$CDM for the DESI+GW+CC+RSD combination to a mild preference for REC model when Type~Ia supernova data are included. In contrast, BIC remains consistently negative because of its stronger penalty for the additional model parameter, and therefore favours $\Lambda$CDM throughout the considered dataset combinations. The BIC preference is strongest for DESI+GW+CC+RSD and DESI+GW+CC+RSD+PantheonPlus, while it becomes substantially weaker for the Union3 combination. The DIC provides a further consistency check, yielding $\Delta{\rm DIC}\simeq0.78$--$3.50$. This again indicates that the improvement in the effective fit of REC model becomes more evident when the supernova observations are incorporated. In particular, the similarity between $\Delta\chi^2$ and $\Delta{\rm DIC}$ for the PantheonPlus and Union3 combinations suggests that the improved fit is not primarily driven by an unfavorable increase in the effective model complexity. Finally, we evaluate the Bayesian preference using the directly computed marginal likelihoods, we obtain $\ln B\simeq-0.73$ to $-1.84$ across the three dataset combinations. Since $\ln B<0$ for every combination, the Bayesian evidence consistently favors REC model over $\Lambda$CDM. Moreover, the preference becomes stronger when either PantheonPlus or Union3 is added, suggesting that the supernova observations play an important role in enhancing the relative statistical performance of the modified model. However, the magnitude of $\ln B$ remains below the level conventionally associated with strong or decisive Bayesian evidence. Taken together, the statistical indicators reveal a consistent but non-decisive preference for REC model.}
\item{\textbf{Evolution of the Cosmological parameters:} To examine the late-time dynamics of the Universe for the REC model, we investigate the redshift evolution of the cosmological parameters, with particular emphasis on the deceleration parameter, effective EoS paramter and the matter density parameter as shown in Fig.~(\ref{f3}). For further analysis of the dynamical properties we also perform statefinder diagnostics for the REC model as shown in Fig.~(\ref{f4}). The evolution of the deceleration parameter exhibits the expected transition from an early decelerating phase to the present accelerated expansion. The transition redshift, determined from the condition $q(z_{\rm tr})=0$, remains remarkably stable across the different dataset combinations, yielding $z_{\rm tr}\simeq0.664$--$0.678$, with typical $1\sigma$ uncertainties of approximately $0.02$. The small variation in $z_{\rm tr}$ demonstrates that the inferred onset of cosmic acceleration is remarkably robust against the inclusion and choice of the Type~Ia supernova compilation. The evolution of the effective equation-of-state parameter, $\omega_{\rm eff}(z)$ exhibits the expected transition from an early matter-dominated, decelerating regime, characterized by $\omega_{\rm eff}(z)\simeq0$ at high redshifts, to a late-time accelerated phase with $\omega_{\rm eff}(z)<-1/3$. The present-day values inferred from the different observational combinations remain tightly clustered within $\omega_{\rm eff}(0)\in[-0.682,-0.677]$, indicating a predominantly quintessence like effective cosmic fluid at the current epoch, without entering the phantom regime ($\omega_{\rm eff}<-1$). Which is well supported by the statefinder analysis, that demonstrates that the REC model evolves in the quintessence region and gradually approaches the $\Lambda$CDM fixed point $(r,s) = (1,0)$ and the de~Sitter attractor $(r,q) = (1,-1)$ at late times. The absence of trajectories entering the phantom regime confirms the dynamical consistency and physical viability of the model. Interestingly, the inclusion of the Type Ia supernova data leads to a slightly less negative present-day $\omega_{\rm eff}$, accompanied by a correspondingly less negative value of the deceleration parameter. This indicates that incorporating the Type Ia supernova data favors a marginally slightly less stronger late-time accelerated expansion in the REC framework. Although the overall evolution of $q(z)$ and $\omega_{eff}(z)$ closely follows the $\Lambda$CDM prediction, the model shows mild departures at both high and low redshifts, with the deviations becoming comparatively more pronounced toward the lower-redshift regime. Nevertheless, these departures remain sufficiently small to preserve the standard late-time accelerated-expansion picture. However, in case of the evolution of the matter density parameter the REC model shows slight deviation only at the higher redshifts. At lower redshiftd the evolution of the density parameter for the REC model becomes similar to that of the $\Lambda$CDM model.
These observations are well consistent with different recent studies as in Refs.~\cite{farooq2013, xu2012, xu2012new, R2,R3,R4,R5,R6,R7,R8}}
\end{enumerate}
The stability of the transition redshift, together with the nearly unchanged constraints on $H_0$, $\Omega_{m0}$, and $r_d$, indicates that the REC model does not significantly modify the overall expansion history relative to $\Lambda$CDM. Rather, its impact is primarily manifested through a controlled modification of the late-time cosmic dynamics. This interpretation is consistent with the statistical analysis, which shows that the REC model provides a modest improvement in the fit to the considered late-time observational datasets without requiring a substantial departure from the standard cosmological expansion history. However, although the model yields a statistically competitive fit, the resulting shifts in $H_0$ and $\sigma_{8,0}$ do not occur in the directions required to simultaneously alleviate the $H_0$ and $S_8$ tensions. Thus, the principal observational implication of the REC framework is a competitive and mildly modified late-time expansion history, rather than a direct resolution of the current cosmological tensions.

\begin{table}[htbp]
  \centering
  \renewcommand{\arraystretch}{1.45} 
  \setlength{\tabcolsep}{7pt} 
  
  \caption{Best-fit values and $1\sigma$ uncertainties of free cosmological parameters for $\Lambda\text{CDM}$ and the REC model across dataset combinations. Here, $D_1$: DESI + GW + CC + RSD, $D_2$: DESI + GW + CC + RSD + PantheonPlus, and $D_3$: DESI + GW + CC + RSD + Union3.}
  \label{tab1}
  \vspace{0.5em}
  
  \begin{tabular}{l l c c c c c}
    \toprule
    \textbf{Dataset} & \textbf{Model} & $H_0\ \mathrm{(km\,s^{-1}\,Mpc^{-1})}$ & $\Omega_{m0}$ & $C\ \mathrm{(Mpc^{2}\,s^{2}\,km^{-2})}$ & $r_d\ \mathrm{(Mpc)}$ & $\sigma_{8,0}$ \\
    \midrule
    \textbf{$D_1$} 
      & $\Lambda\text{CDM}$ & $68.92_{-1.48}^{+1.43}$ & $0.2944_{-0.0080}^{+0.0083}$ & --- & $147.57_{-2.97}^{+3.21}$ & $0.7962_{-0.0247}^{+0.0244}$ \\
      & REC         & $68.42_{-1.41}^{+1.44}$ & $0.2911_{-0.0081}^{+0.0083}$ & $0.9930_{-0.0979}^{+0.1062}$ & $147.36_{-2.97}^{+3.11}$ & $0.8152_{-0.0259}^{+0.0259}$ \\
    \midrule
    \textbf{$D_2$} 
      & $\Lambda\text{CDM}$ & $68.65_{-1.44}^{+1.42}$ & $0.3011_{-0.0072}^{+0.0073}$ & --- & $147.42_{-2.95}^{+3.17}$ & $0.7942_{-0.0251}^{+0.0250}$ \\
      & REC         & $68.34_{-1.39}^{+1.39}$ & $0.2945_{-0.0075}^{+0.0076}$ & $1.0007_{-0.1012}^{+0.1035}$ & $147.21_{-2.94}^{+2.98}$ & $0.8129_{-0.0248}^{+0.0245}$ \\
    \midrule
    \textbf{$D_3$} 
      & $\Lambda\text{CDM}$ & $68.71_{-1.45}^{+1.41}$ & $0.3005_{-0.0076}^{+0.0081}$ & --- & $147.37_{-2.96}^{+3.15}$ & $0.7948_{-0.0248}^{+0.0250}$ \\
      & REC         & $68.24_{-1.41}^{+1.39}$ & $0.2953_{-0.0079}^{+0.0080}$ & $0.9995_{-0.1027}^{+0.1034}$ & $147.32_{-2.92}^{+3.06}$ & $0.8125_{-0.0261}^{+0.0252}$ \\
    \bottomrule
  \end{tabular}
\end{table}

\begin{table}[htbp]
  \centering
  \renewcommand{\arraystretch}{1.45} 
  \setlength{\tabcolsep}{8pt} 

  \caption{Statistical model comparison between $\Lambda\text{CDM}$ and REC across different dataset combinations. Differences are defined as $\Delta X = X_{\Lambda\mathrm{CDM}} - X_{\mathrm{REC}}$ (where $\Delta X > 0$ favors REC) and $\ln B = \ln \mathcal{Z}_{\Lambda\mathrm{CDM}} - \ln \mathcal{Z}_{\mathrm{REC}}$ (where $\ln B < 0$ favors REC). Here, $D_1$: DESI + GW + CC + RSD, $D_2$: DESI + GW + CC + RSD + PantheonPlus, and $D_3$: DESI + GW + CC + RSD + Union3.}
  \label{tab3}
  \vspace{0.5em}
  
  \begin{tabular}{l c c c c c}
    \toprule
    \textbf{Dataset} & $\Delta\chi^2_{\min}$ & $\Delta\mathrm{AIC}$ & $\Delta\mathrm{BIC}$ & $\Delta\mathrm{DIC}$ & $\ln B$ \\
    \midrule
    \textbf{$D_1$} 
      & $+0.8593$ 
      & $-1.1407$ 
      & $-4.1842$ 
      & $+0.7844$ 
      & $-0.7266 \pm 0.2205$ \\
    \midrule
    \textbf{$D_2$} 
      & $+3.4799$ 
      & $+1.4799$ 
      & $-3.9846$ 
      & $+3.4842$ 
      & $-1.6304 \pm 0.2194$ \\
    \midrule
    \textbf{$D_3$} 
      & $+3.5185$ 
      & $+1.5185$ 
      & $-1.6577$ 
      & $+3.4969$ 
      & $-1.8360 \pm 0.2216$ \\
    \bottomrule
  \end{tabular}
\end{table}

\begin{table}[htbp]
  \centering
  \renewcommand{\arraystretch}{1.45} 
  \setlength{\tabcolsep}{6pt} 

  \caption{Derived cosmological quantities at $z=0$ and transition redshift $z_{\mathrm{tr}}$ ($1\sigma$ uncertainties) across dataset combinations. Here, $D_1$: DESI + GW + CC + RSD, $D_2$: DESI + GW + CC + RSD + PantheonPlus, and $D_3$: DESI + GW + CC + RSD + Union3.}
  \label{tab2}
  \vspace{0.5em}
  
  \begin{tabular}{l c c c c c}
    \toprule
    \textbf{Dataset} & $q(0)$ & $w_{\mathrm{eff}}(0)$ & $f\sigma_8(0)$ & $\Omega_{\mathrm{DE}}(0)$ & $z_{\mathrm{tr}}$ \\
    \midrule
    \textbf{$D_1$} 
      & $-0.5231_{-0.0123}^{+0.0134}$ 
      & $-0.6821_{-0.0085}^{+0.0086}$ 
      & $+0.4129_{-0.0130}^{+0.0132}$ 
      & $+0.7088_{-0.0080}^{+0.0078}$ 
      & $0.6776_{-0.0212}^{+0.0235}$ \\
    \midrule
    \textbf{$D_2$} 
      & $-0.5175_{-0.0121}^{+0.0119}$ 
      & $-0.6784_{-0.0076}^{+0.0083}$ 
      & $+0.4153_{-0.0134}^{+0.0140}$ 
      & $+0.7053_{-0.0073}^{+0.0073}$ 
      & $0.6678_{-0.0207}^{+0.0199}$ \\
    \midrule
    \textbf{$D_3$} 
      & $-0.5155_{-0.0124}^{+0.0128}$ 
      & $-0.6770_{-0.0081}^{+0.0080}$ 
      & $+0.4159_{-0.0140}^{+0.0138}$ 
      & $+0.7040_{-0.0075}^{+0.0082}$ 
      & $0.6643_{-0.0219}^{+0.0230}$ \\
    \bottomrule
  \end{tabular}
\end{table}

\begin{table}[htbp]
\centering
\caption{Derived constraints on the Rényi parameter $\lambda$ with 1$\sigma$ confidence level. The constraints are obtained from Eq.~(\ref{E5x}) using the posterior distributions of $H_0$, $\Omega_{m0}$, and $C$. In the physical-unit formulation, we adopt $(8\pi G)^{-1/2}\simeq 2.4\times10^{18}\,\mathrm{GeV}$ and express $H_0$ and $C$ in GeV, with all dimensional quantities entering the logarithm consistently converted to natural units.}
\label{tab4}
\renewcommand{\arraystretch}{1.5}
\setlength{\tabcolsep}{12pt}
\begin{tabular}{lc}
\hline
\textbf{Dataset combination} &
\textbf{$\lambda$} \\
\hline
DESI + GW + CC + RSD
&
$(3.769^{+0.166}_{-0.154})\times10^{-124}$ \\[6pt]

DESI + GW + CC + RSD + PantheonPlus
&
$(3.743^{+0.152}_{-0.153})\times10^{-124}$ \\[6pt]

DESI + GW + CC + RSD + Union3
&
$(3.727^{+0.155}_{-0.153})\times10^{-124}$ \\
\hline
\end{tabular}
\end{table}

\section{Conclusion}\label{s5}
In this work, we investigated the cosmological implications of Rényi entropic cosmology (REC), in which deviations from the standard cosmological dynamics arise from non-extensive corrections to the horizon entropy governed by the Rényi parameter $\lambda$. The model was confronted with three combinations of late-time observations, namely DESI+GW+CC+RSD, DESI+GW+CC+RSD+PantheonPlus, and DESI+GW+CC+RSD+Union3, using a comprehensive MCMC analysis. The resulting constraints show that the model parameters remain remarkably stable across the different dataset combinations, indicating that the inferred cosmological evolution is largely insensitive to the choice of Type~Ia supernova compilation. In particular, REC yields $H_0\simeq68.24$--$68.42$ $km\,s^{-1}\,Mpc^{-1}$, $\Omega_{m0}\simeq0.291$--$0.295$, $C\simeq0.99$--$1.00$, and $r_d\simeq147.21$--$147.36$~$\mathrm{Mpc}$. Its inferred Hubble constant remains close to the $\Lambda$CDM value and well below the local distance-ladder determination, while the preferred clustering amplitude, $\sigma_{8,0}\simeq0.813$--$0.815$, is systematically higher than that of $\Lambda$CDM. From the perspective of cosmological tensions, REC model does not significantly alleviate either the $H_0$ or $S_8$ tension. This behavior is consistent with most of the recent late-time analyses, which generally find that dynamical dark-energy extensions can modify the inferred expansion history without necessarily resolving the persistent $S_8$ and $H_0$ discrepancy \cite{DJ1,farooq2013, xu2012, xu2012new, R2,R3}.\\
The statistical comparison with $\Lambda$CDM shows that REC model consistently improves the minimum $\chi^2$, with $\Delta\chi^2\simeq0.86$--$3.52$. The improvement becomes more evident when the Type~Ia supernova data are included. The logarithmic Bayes factor ($\ln B$) favours REC model for all dataset combinations, but it's magnitude remains below the level associated with strong or decisive evidence. The AIC provides only mild support, whereas BIC remains largely in favour of the $\Lambda$CDM because of its stronger complexity penalty that it puts on the REC model. Therefore, the present analysis indicates that REC model is a statistically competitive extension of $\Lambda$CDM, with the strongest relative improvement emerging from the joint late-time datasets, but the current observations do not provide decisive evidence for a departure from the standard cosmological model. The reconstructed cosmic dynamics further demonstrate the consistency of the model with the observed transition to accelerated expansion. The deceleration parameter exhibits a robust transition from an early decelerating epoch to late-time acceleration at $z_{\rm tr}\simeq0.664$--$0.678$~\cite{farooq2013, xu2012, xu2012new}. Similarly, the effective equation-of-state parameter evolves from a matter-like regime, $\omega_{\rm eff}\simeq0$, at high redshift to a quintessence-like late-time regime, with $\omega_{\rm eff}(0)\simeq-0.682$ to $-0.677$, without crossing the phantom boundary. The statefinder trajectories remain in the quintessence region and approach the $\Lambda$CDM fixed point $(r,s)=(1,0)$ and the de~Sitter attractor $(r,q)=(1,-1)$ at late times. The matter-density evolution remains close to that of $\Lambda$CDM at low redshifts, with only mild deviations emerging mainly at higher redshifts.\\
The novelty of the present work does not lie in the construction of modified Friedmann equations from entropic corrections, which has already been extensively explored for several generalized entropy prescriptions~\cite{R2,R3,R4,R5,R6,R7,R8}. Rather, we provide a dedicated and updated observational assessment of the Rényi entropic cosmology (REC) framework using complementary late-time probes, including DESI BAO, gravitational-wave standard sirens, cosmic chronometers, RSD, PantheonPlus, and Union3, together with a multi-criterion statistical comparison and a unified analysis of its cosmological dynamics. Different generalized-entropy models ~\cite{R2,R3,R4,R5,R6,R7,R8} can generate similar late-time $\Lambda$CDM-like backgrounds, but they differ in the extent to which they shift $H_0$, the clustering amplitude, and the statistical preference for departures from $\Lambda$CDM. REC, in the present analysis, remains close to $\Lambda$CDM with a a fit to the observational data that is competitive but not decisive, at the same time does not resolve either the $H_0$ or $S_8$ tensions. Additionally, as shown in Table~(\ref{tab4}) the Rényi parameter is constrained as $\lambda \sim (3.7-3.8) \times10^{-124}$. Interestingly, these values are compatible with the parameter range discussed in Ref.~\cite{n1}, which investigated the viability of Rényi entropic cosmology in the context of Big Bang nucleosynthesis and baryogenesis. Thus, the late-time observational constraints obtained here remain consistent with the independent early-Universe viability considerations reported in the literature.\\
Overall, the REC framework provides a viable and statistically competitive description of the late-time Universe, while preserving an expansion history close to that of $\Lambda$CDM. Its main distinguishing feature is therefore not a dramatic departure from standard cosmology, but a controlled modification of the late-time dynamics arising from the underlying entropic correction. Future high-precision observations of cosmic expansion and structure growth will be essential for testing these small deviations and determining whether they can provide a meaningful observational distinction between REC and the $\Lambda$CDM paradigm.

\section*{Data Availability}
All data used in this study are cited
in the references and were obtained from publicly available sources.

.


\begin{thebibliography}{200}


\bibitem{riess1998} 
Riess, A. G., et al. (1998). Observational Evidence from Supernovae for an Accelerating Universe and a Cosmological Constant. \textit{Astronomical Journal}, \textit{116}(3), 1009–1038. \href{https://doi.org/10.1086/300499}{https://doi.org/10.1086/300499}

\bibitem{perlmutter1999} 
Perlmutter, S., et al. (1999). Measurements of $\Omega$ and $\Lambda$ from 42 High-Redshift Supernovae. \textit{Astrophysical Journal}, \textit{517}(2), 565–586. \href{https://doi.org/10.1086/307221}{https://doi.org/10.1086/307221}

\bibitem{bennett2003} 
Bennett, C. L., et al. (2003). First-Year Wilkinson Microwave Anisotropy Probe (WMAP) Observations: Preliminary Maps and Basic Results. \textit{Astrophysical Journal Supplement Series}, \textit{148}, 1–27. \href{https://doi.org/10.1086/377253}{https://doi.org/10.1086/377253}

\bibitem{spergel2003} 
Spergel, D. N., et al. (2003). First-Year Wilkinson Microwave Anisotropy Probe (WMAP) Observations: Determination of Cosmological Parameters. \textit{Astrophysical Journal Supplement Series}, \textit{148}, 175–194. \href{https://doi.org/10.1086/377226}{https://doi.org/10.1086/377226}

\bibitem{eisenstein2005} 
Eisenstein, D. J., et al. (2005). Detection of the Baryon Acoustic Peak in the Large-Scale Correlation Function of SDSS Luminous Red Galaxies. \textit{Astrophysical Journal}, \textit{633}(2), 560–574. \href{https://doi.org/10.1086/466512}{https://doi.org/10.1086/466512}

\bibitem{percival2010} 
Percival, W. J., et al. (2010). Baryon Acoustic Oscillations in the Sloan Digital Sky Survey Data Release 7 Galaxy Sample. \textit{Monthly Notices of the Royal Astronomical Society}, \textit{401}(4), 2148–2168. \href{https://doi.org/10.1111/j.1365-2966.2009.15812.x}{https://doi.org/10.1111/j.1365-2966.2009.15812.x}

\bibitem{weinberg1989} 
Weinberg, S. (1989). The Cosmological Constant Problem. \textit{Reviews of Modern Physics}, \textit{61}(1), 1–23. \href{https://doi.org/10.1103/RevModPhys.61.1}{https://doi.org/10.1103/RevModPhys.61.1}

\bibitem{Martin2012}
J.~Martin, ``Everything You Always Wanted To Know About The Cosmological Constant Problem (But Were Afraid To Ask),'' \textit{Comptes Rendus Physique} \textbf{13}, 566 (2012), \href{https://arxiv.org/abs/1205.3365}{arXiv:1205.3365 [astro-ph.CO]}.


\bibitem{carroll1992} 
Carroll, S. M., Press, W. H., \& Turner, E. L. (1992). The Cosmological Constant. \textit{Annual Review of Astronomy and Astrophysics}, \textit{30}, 499–542. \href{https://doi.org/10.1146/annurev.aa.30.090192.002435}{https://doi.org/10.1146/annurev.aa.30.090192.002435}

\bibitem{sahni2000} 
Sahni, V., \& Starobinsky, A. A. (2000). The Case for a Positive Cosmological $\Lambda$-Term. \textit{International Journal of Modern Physics D}, \textit{9}(4), 373–443. \href{https://doi.org/10.1142/S0218271800000542}{https://doi.org/10.1142/S0218271800000542}

\bibitem{ArmendarizPicon2000}
Armendariz-Picon, C., Mukhanov, V., \& Steinhardt, P. J. (2000). Dynamical solution to the problem of a small cosmological constant and late-time cosmic acceleration. \textit{Phys. Rev. Lett.}, \textit{85}(21), 4438–4441. \href{https://doi.org/10.1103/PhysRevLett.85.4438}{https://doi.org/10.1103/PhysRevLett.85.4438}

\bibitem{Sen2002}
Sen, A. (2002). Rolling tachyon. \textit{Journal of High Energy Physics}, \textit{2002}(04), 048. \href{https://doi.org/10.1088/1126-6708/2002/04/048}{https://doi.org/10.1088/1126-6708/2002/04/048}

\bibitem{Khoury2004}
Khoury, J., \& Weltman, A. (2004). Chameleon fields: Awaiting surprises for tests of gravity in space. \textit{Phys. Rev. Lett.}, \textit{93}(17), 171104. \href{https://doi.org/10.1103/PhysRevLett.93.171104}{https://doi.org/10.1103/PhysRevLett.93.171104}

\bibitem{Kamenshchik2001}
Kamenshchik, A., Moschella, U., \& Pasquier, V. (2001). An alternative to quintessence. \textit{Physics Letters B}, \textit{511}(2-4), 265–268. \href{https://doi.org/10.1016/S0370-2693(01)00571-8}{https://doi.org/10.1016/S0370-2693(01)00571-8}

\bibitem{Bento2002}
Bento, M. C., Bertolami, O., \& Sen, A. A. (2002). Generalized Chaplygin gas, accelerated expansion, and dark energy-matter unification. \textit{Phys. Rev. D}, \textit{66}(4), 043507. \href{https://doi.org/10.1103/PhysRevD.66.043507}{https://doi.org/10.1103/PhysRevD.66.043507}


\bibitem{bardeen1973}
J.~M.~Bardeen, B.~Carter and S.~W.~Hawking,
The four laws of black hole mechanics,
\textit{Commun. Math. Phys.} \textbf{31}, 161 (1973).
\href{https://doi.org/10.1007/BF01645742}{https://doi.org/10.1007/BF01645742}

\bibitem{bekenstein1973}
J.~D.~Bekenstein,
Black holes and entropy,
\textit{Phys. Rev. D} \textbf{7}, 2333 (1973).
\href{https://doi.org/10.1103/PhysRevD.7.2333}{https://doi.org/10.1103/PhysRevD.7.2333}

\bibitem{hawking1975}
S.~W.~Hawking,
Particle creation by black holes,
\textit{Commun. Math. Phys.} \textbf{43}, 199 (1975).
\href{https://doi.org/10.1007/BF02345020}{https://doi.org/10.1007/BF02345020}

\bibitem{tHooft1993}
G.~'t~Hooft,
Dimensional reduction in quantum gravity,
in \textit{Salamfestschrift}, World Scientific (1993).
\href{https://arxiv.org/abs/gr-qc/9310026}{https://arxiv.org/abs/gr-qc/9310026}

\bibitem{susskind1995}
L.~Susskind,
The world as a hologram,
\textit{J. Math. Phys.} \textbf{36}, 6377 (1995).
\href{https://doi.org/10.1063/1.531249}{https://doi.org/10.1063/1.531249}

\bibitem{maldacena1999}
J.~M.~Maldacena,
The large $N$ limit of superconformal field theories and supergravity,
\textit{Adv. Theor. Math. Phys.} \textbf{2}, 231 (1998).
\href{https://doi.org/10.1023/A:1026654312961}{https://doi.org/10.1023/A:1026654312961}

\bibitem{almheiri2013}
A.~Almheiri, D.~Marolf, J.~Polchinski and J.~Sully,
Black holes: Complementarity or firewalls?,
\textit{JHEP} \textbf{02}, 062 (2013).
\href{https://doi.org/10.1007/JHEP02(2013)062}{https://doi.org/10.1007/JHEP02(2013)062}

\bibitem{jacobson1995}
T.~Jacobson,
Thermodynamics of spacetime: The Einstein equation of state,
\textit{Phys. Rev. Lett.} \textbf{75}, 1260 (1995).
\href{https://doi.org/10.1103/PhysRevLett.75.1260}{https://doi.org/10.1103/PhysRevLett.75.1260}

\bibitem{cai2005}
R.~G.~Cai and S.~P.~Kim,
First law of thermodynamics and Friedmann equations of FRW universe,
\textit{JHEP} \textbf{02}, 050 (2005).
\href{https://doi.org/10.1088/1126-6708/2005/02/050}{https://doi.org/10.1088/1126-6708/2005/02/050}

\bibitem{akbar2006}
M.~Akbar and R.~G.~Cai,
Thermodynamic behavior of Friedmann equations at apparent horizon,
\textit{Phys. Rev. D} \textbf{75}, 084003 (2007).
\href{https://doi.org/10.1103/PhysRevD.75.084003}{https://doi.org/10.1103/PhysRevD.75.084003}

\bibitem{sheykhi2018}
A.~Sheykhi,
Modified Friedmann equations from generalized entropy,
\textit{Phys. Lett. B} \textbf{785}, 118 (2018).
\href{https://doi.org/10.1016/j.physletb.2018.08.018}{https://doi.org/10.1016/j.physletb.2018.08.018}

\bibitem{nojiri2017}
S.~Nojiri, S.~D.~Odintsov and V.~K.~Oikonomou,
Modified gravity theories on a nutshell,
\textit{Phys. Rept.} \textbf{692}, 1 (2017).
\href{https://doi.org/10.1016/j.physrep.2017.06.001}{https://doi.org/10.1016/j.physrep.2017.06.001}

\bibitem{tsallis1988}
C.~Tsallis,
Possible generalization of Boltzmann--Gibbs statistics,
\textit{J. Stat. Phys.} \textbf{52}, 479 (1988).
\href{https://doi.org/10.1007/BF01016429}{https://doi.org/10.1007/BF01016429}

\bibitem{renyi1961}
A.~Rényi,
On measures of entropy and information,
in \textit{Proceedings of the Fourth Berkeley Symposium on Mathematical Statistics and Probability} (1961).

\bibitem{barrow2020}
J.~D.~Barrow,
The area of a rough black hole,
\textit{Phys. Lett. B} \textbf{808}, 135643 (2020).
\href{https://doi.org/10.1016/j.physletb.2020.135643}{https://doi.org/10.1016/j.physletb.2020.135643}

\bibitem{kaniadakis2002}
G.~Kaniadakis,
Statistical mechanics in the context of special relativity,
\textit{Phys. Rev. E} \textbf{66}, 056125 (2002).
\href{https://doi.org/10.1103/PhysRevE.66.056125}{https://doi.org/10.1103/PhysRevE.66.056125}

\bibitem{sharma1975}
B.~D.~Sharma and D.~P.~Mittal,
New non-additive measures of entropy,
\textit{J. Math. Sci.} \textbf{10}, 28 (1975).

\bibitem{rovelli1996}
C.~Rovelli,
Black hole entropy from loop quantum gravity,
\textit{Phys. Rev. Lett.} \textbf{77}, 3288 (1996).
\href{https://doi.org/10.1103/PhysRevLett.77.3288}{https://doi.org/10.1103/PhysRevLett.77.3288}

\bibitem{x13} Das, S., S. Shankaranarayanan, \& Sur, S. (2008). Power-law corrections to entanglement entropy of horizons. Physical Review. D. Particles, Fields, Gravitation, and Cosmology/Physical Review. D, Particles, Fields, Gravitation, and Cosmology, 77(6). \href{https://doi.org/10.1103/physrevd.77.064013}{doi:10.1103/physrevd.77.064013}.
‌
\bibitem{x14} Radicella, N., \& Pavón, D. (2010). The generalized second law in universes with quantum corrected entropy relations. Physics Letters B, 691(3), 121–126. \href{https://doi.org/10.1016/j.physletb.2010.06.019}{doi:10.1016/j.physletb.2010.06.019}
‌
\bibitem{x16} Mann, R. B., \& Solodukhin, S. N. (1997). Quantum scalar field on a three-dimensional (BTZ) black hole instanton: Heat kernel, effective action, and thermodynamics. Physical Review. D. Particles, Fields, Gravitation, and Cosmology/Physical Review. D. Particles and Fields, 55(6), 3622–3632. \href{https://doi.org/10.1103/physrevd.55.3622}{doi:10.1103/physrevd.55.3622}
‌
\bibitem{x21} Banerjee, R., \& Majhi, B. R. (2008). Quantum tunneling and back reaction. Physics Letters B, 662(1), 62–65. \href{https://doi.org/10.1016/j.physletb.2008.02.044}{doi:10.1016/j.physletb.2008.02.044}
‌
\bibitem{x22} Shannon, C. E. (1948). A Mathematical Theory of Communication. Bell System Technical Journal, 27(3), 379–423. \href{https://doi.org/10.1002/j.1538-7305.1948.tb01338.x}{doi:10.1002/j.1538-7305.1948.tb01338.x}

\bibitem{n1} Sheykhi, A., \& Sooraki, A. S. (2025). Constraints on Renyi Entropy through Primordial Big-Bang Nucleosynthesis and Baryogenesis. \href{https://arxiv.org/pdf/2507.14250}{2507.14250}

\bibitem{n2} Fazlollahi, H. R. (2023). Rényi entropy correction to expanding universe. The European Physical Journal C, 83(1). \href{https://doi.org/10.1140/epjc/s10052-023-11183-w}{doi:10.1140/epjc/s10052-023-11183-w}

\bibitem{n3} Ghaffari, S., Ziaie, A. H., Bezerra, V. B., \& Moradpour, H. (2019). Inflation in the Rényi cosmology. Modern Physics Letters A, 35(01), 1950341. \href{https://doi.org/10.1142/s0217732319503413}{doi:10.1142/s0217732319503413}
‌ 
\bibitem{n4} Golanbari, T., Saaidi, K., \& Karimi, P. (2020). Renyi entropy and the holographic dark energy in flat space time. \href{ https://arxiv.org/abs/2002.04097}{2002.04097} 

\bibitem{n5} Pandey, B. (2021). Renyi entropy as a measure of cosmic homogeneity. Journal of Cosmology and Astroparticle Physics, 2021(02), 023–023. \href{https://doi.org/10.1088/1475-7516/2021/02/023}{doi:10.1088/1475-7516/2021/02/023}


‌\bibitem{ref90} Abdul Karim, M., Aguilar, J., Ahlen, S., Alam, S., Allen, L., Prieto, C. A., Brieden, S. (2025). DESI DR2 results. II. Measurements of baryon acoustic oscillations and cosmological constraints. Physical Review D, 112(8). \href{https://doi.org/10.1103/tr6y-kpc6}{https://doi.org/10.1103/tr6y-kpc6}

\bibitem{refk58}
A. N. Ormondroyd, W. J. Handley, M. P. Hobson, and A. N. Lasenby. \textit{Comparison of Dynamical Dark Energy with \(\Lambda\)CDM in Light of DESI DR2}. \href{https://arxiv.org/abs/2503.17342}{arXiv:2503.17342}


\bibitem{refk65}
E. Chaussidon \textit{et al.}. (2025).  
\textit{Early Time Solution as an Alternative to the Late Time Evolving Dark Energy with DESI DR2 BAO}.  
\href{https://arxiv.org/abs/2503.24343}{[arXiv:2503.24343]}

\bibitem{refk68}
Wolf, W. J., García-García, C., Anton, T., et al. (2025). Assessing Cosmological Evidence for Nonminimal Coupling. \textit{Phys. Rev. Lett.}, \textit{135}(8), 081001. \href{https://doi.org/10.1103/jysf-k72m}{https://doi.org/10.1103/jysf-k72m}
\href{https://arxiv.org/abs/2504.07679}{[arXiv:2504.07679]}

\bibitem{refk69}
A. Paliathanasis. (2025).  
\textit{Dark Energy within the Generalized Uncertainty Principle in Light of DESI DR2}.  
\href{https://arxiv.org/abs/2503.20896}{[arXiv:2503.20896]}

\bibitem{g1} Rich Abbott et. al. (2021). Open data from the first and second observing runs of Advanced LIGO and Advanced Virgo. SoftwareX, 13, 100658. \href{https://doi.org/10.1016/j.softx.2021.100658}{doi:10.1016/j.softx.2021.100658}

\bibitem{g2} Rich Abbott et. al. GWTC-3: Compact Binary Coalescences Observed by LIGO and Virgo during the Second Part of the Third Observing Run. Physical Review X, 13(4). \href{https://doi.org/10.1103/physrevx.13.041039}{doi:10.1103/physrevx.13.041039}.

\bibitem{PK34} Jimenez, R., Verde, L., Treu, T., \& Stern, D. (2003). \textit{Astrophysical Journal}, \textit{593}, 622. \href{https://arxiv.org/abs/astro-ph/0302560}{arXiv:astro-ph/0302560}

\bibitem{PK35} Moresco, M., et al. (2012). \textit{Journal of Cosmology and Astroparticle Physics}, \textit{1208}, 006. \href{https://arxiv.org/abs/1201.3609}{arXiv:1201.3609 [astro-ph.CO]}

\bibitem{PK36} Moresco, M. (2015). \textit{Monthly Notices of the Royal Astronomical Society}, \textit{450}, L16. \href{https://arxiv.org/abs/1503.01116}{arXiv:1503.01116 [astro-ph.CO]}

\bibitem{PK37} Moresco, M., Pozzetti, L., Cimatti, A., Jimenez, R., Maraston, C., Verde, L., Thomas, D., Citro, A., Tojeiro, R., \& Wilkinson, D. (2016). \textit{Journal of Cosmology and Astroparticle Physics}, \textit{1605}, 014. \href{https://arxiv.org/abs/1601.01701}{arXiv:1601.01701 [astro-ph.CO]}

\bibitem{PK41} Alestas, G., Kazantzidis, L., \& Nesseris, S. (2022). \textit{Physical Review D}, \textit{106}, 103519. \href{https://arxiv.org/abs/2209.12799}{arXiv:2209.12799 [astro-ph.CO]}

\bibitem{PK45} Brout D, Scolnic D, Popovic B, et al. The Pantheon+ Analysis: Cosmological constraints. The Astrophysical Journal. 2022;938(2):110. \href{https://doi.org/10.3847/1538-4357/ac8e04}{doi:10.3847/1538-4357/ac8e04}

\bibitem{PK48} Rubin, D., Aldering, G., Betoule, M., Fruchter, A., Huang, X., Kim, A. G., Lidman, C., Linder, E., Perlmutter, S., Ruiz-Lapuente, P., et al. (2025). Union through UNITY: Cosmology with 2000 SNe Using a Unified Bayesian Framework. \textit{Astrophysical Journal}, \textit{986}, 231. \href{https://doi.org/10.3847/1538-4357/adc0a5}{https://doi.org/10.3847/1538-4357/adc0a5}



\bibitem{PK38} Kavya, N. S., Mishra, S. S., \& Sahoo, P. K. (2025). \textit{Scientific Reports}, \textit{15}, 36504.

\bibitem{PK39} Moresco, M., Jimenez, R., Verde, L., Cimatti, A., \& Pozzetti, L. (2020). \textit{Astrophysical Journal}, \textit{898}, 82. \href{https://arxiv.org/abs/2003.07362}{arXiv:2003.07362 [astro-ph.GA]}
‌


\bibitem{AK1} Yadav, M., Dixit, A., Barak, M. S., \& Pradhan, A. (2026). Constraints on spatial curvature and dark energy dynamics in the wCDM model from DESI DR1 and DR2. Journal of High Energy Astrophysics, 50, 100514. 

\bibitem{PKx} Kavya, N. S., Swagat Mishra, S., \& Sahoo, P. K. (2025). $f(Q)$ gravity as a possible resolution of the $H_0$ and $S_8$ tensions with DESI DR2. Scientific Reports, 15(1). \href{https://doi.org/10.1038/s41598-025-23502-0}{https://doi.org/10.1038/s41598-025-23502-0}
‌
‌


\bibitem{PK40} Macaulay, E., Wehus, I. K., \& Eriksen, H. K. (2013). \textit{Physical Review Letters}, \textit{111}, 161301. \href{https://arxiv.org/abs/1303.6583}{arXiv:1303.6583 [astro-ph.CO]}



\bibitem{PK46} Scolnic D, Brout D, Carr A, et al. The Pantheon+ analysis: the full data set and light-curve release. The Astrophysical Journal. 2022;938(2):113. \href{https://doi.org/10.3847/1538-4357/ac8b7a}{doi:10.3847/1538-4357/ac8b7a}




\bibitem{PKx1} Sai Swagat, M., \& Sahoo, P. K. (2026). Alleviating Cosmological Tensions in the Topological Dark Energy Framework. Progress of Theoretical and Experimental Physics, 2026(6). \href{https://academic.oup.com/ptep/article/2026/6/063E01/8667130}{DOI: 10.1093/ptep/ptag080}

\bibitem{DJ1} Chaudhary, H., Jyoti Gogoi, D., Capozziello, S., \& Mustafa, G. (2026). Constraints on the Modified Emergent Dark Energy Model Using DESI DR2 BAO. Journal of High Energy Astrophysics, 100716. \href{https://doi.org/10.1016/j.jheap.2026.100716}{DOI:10.1016/j.jheap.2026.100716}

\bibitem{PK47} Mateus S., et al. Challenging the $\Lambda$CDM model: 5$\sigma$ evidence for a dynamical dark energy late-time transition. Physical Review D/Physical Review D. July 2025. \href{https://doi.org/10.1103/n86r-sjgm}{doi:10.1103/n86r-sjgm}



\bibitem{PK49} Mishra, S. S., Kavya, N. S., Sahoo, P. K., \& Harko, T. (2025). Padé cosmography and its insights into teleparallel gravity. \textit{Monthly Notices of the Royal Astronomical Society}, \textit{543}, 2816. \href{https://doi.org/10.1093/mnras/staf1492}{https://doi.org/10.1093/mnras/staf1492}


\bibitem{S11}Abbott TMC, Aguena M, Alarcon A, et al. Dark Energy Survey Year 3 results: Cosmological constraints from galaxy clustering and weak lensing. Physical Review D/Physical Review D. 2022;105(2). \href{https://doi.org/10.1103/physrevd.105.023520}{doi:10.1103/physrevd.105.023520}

\bibitem{S12} Asgari M, Lin CA, Joachimi B, et al. KiDS-1000 cosmology: Cosmic shear constraints and comparison between two point statistics. Astronomy and Astrophysics. 2020;645:A104. \href{https://doi.org/10.1051/0004-6361/202039070}{doi:10.1051/0004-6361/202039070}

\bibitem{verde2019tensions} Verde, L., Treu, T., \& Riess, A. G. (2019). Tensions between the early and late Universe. \textit{Nature Astronomy}, \textit{3}, 891–895. \href{https://arxiv.org/abs/1907.10625}{arXiv:1907.10625}

\bibitem{S13} Luciano GG, Paliathanasis A. Modified cosmology through generalized mass-to-horizon entropy: Observational constraints from DESI DR2 BAO data. Physics Letters B. 2025;870:139954. \href{https://doi.org/10.1016/j.physletb.2025.139954}{doi:10.1016/j.physletb.2025.139954}

\bibitem{S13} Riess AG, Yuan W, Macri LM, et al. A Comprehensive Measurement of the Local Value of the Hubble Constant with 1 $km\,s^{-1}\,Mpc^{-1}$ Uncertainty from the Hubble Space Telescope and the SH0ES Team. The Astrophysical Journal Letters. 2022;934(1):L7. \href{doi:10.3847/2041-8213/ac5c5b}{doi:10.3847/2041-8213/ac5c5b}

\bibitem{S14} Planck Collaboration, Aghanim, N., Akrami, Y., et al. (2020). Planck 2018 results. VI. Cosmological parameters. \textit{Astronomy \& Astrophysics}, \textit{641}, A6. \href{https://arxiv.org/abs/1807.06209}{arXiv:1807.06209}

\bibitem{farooq2013} Farooq, O., \& Ratra, B. (2013). Hubble parameter measurement constraints on the cosmological deceleration–acceleration transition redshift. \textit{The Astrophysical Journal Letters}, \textit{766}(1), L7. \href{https://arxiv.org/abs/1301.5243}{arXiv:1301.5243}

\bibitem{xu2012} Zheng, W., Li, H., Xia, J.-Q., Wan, Y.-P., Li, S.-Y., \& Li, M. (2014). Constraints on cosmological models from Hubble parameters measurements. \textit{International Journal of Modern Physics D}, \textbf{23}(05), 1450051. \href{https://doi.org/10.1142/s0218271814500515}{https://doi.org/10.1142/s0218271814500515}.

\bibitem{xu2012new} Kumar, D., Jain, D., Mahajan, S., Mukherjee, A., \& Rana, A. (2023). Constraints on the transition redshift using Hubble phase space portrait. \textit{International Journal of Modern Physics D}, \textbf{23}(06). \href{https://doi.org/10.1142/s0218271823500396}{https://doi.org/10.1142/s0218271823500396}

\bibitem{R2} Nojiri, S., Odintsov, S. D., \& Paul, T. (2022). Early and late universe holographic cosmology from a new generalized entropy. \textit{Physics Letters B}, \textit{831}, 137189. \href{https://doi.org/10.1016/j.physletb.2022.137189}{https://doi.org/10.1016/j.physletb.2022.137189}

\bibitem{R3} Giuseppe, G., L., Paliathanasis, A. (2025). Modified cosmology through generalized mass-to-horizon entropy:
Observational constraints from DESI DR2 BAO data. \textit{Physics Letters B}. \href{https://www.sciencedirect.com/science/article/pii/S0370269325007129}{https://www.sciencedirect.com/science/article/pii/S0370269325007129}

\bibitem{R4} Mendoza-Martínez, M. L., Cervantes-Contreras, A., Trejo-Alonso, J. J., \& Hernandez-Almada, A. (2024). Constraints on Tsallis cosmology using recent low and high redshift measurements. \textit{The European Physical Journal C}, \textit{84}, 717. \href{https://doi.org/10.1140/epjc/s10052-024-13099-5}{https://doi.org/10.1140/epjc/s10052-024-13099-5}

\bibitem{R5} Lymperis, A., \& Saridakis, E. N. (2018). Modified cosmology through nonextensive horizon thermodynamics. \textit{The European Physical Journal C}, \textit{78}, 993. \href{https://doi.org/10.1140/epjc/s10052-018-6480-y}{https://doi.org/10.1140/epjc/s10052-018-6480-y}

\bibitem{R6} Lymperis, A., Basilakos, S., \& Saridakis, E. N. (2021). Modified cosmology through Kaniadakis horizon entropy. \textit{The European Physical Journal C}, \textit{81}, 1037. \href{https://doi.org/10.1140/epjc/s10052-021-09852-9}{https://doi.org/10.1140/epjc/s10052-021-09852-9}

\bibitem{R7} Basilakos, S., Lymperis, A., Petronikolou, M., \& Saridakis, E. N. (2025). Modified cosmology through spacetime thermodynamics and generalized mass-to-horizon entropy. \textit{The European Physical Journal C}, \textit{85}, 1244. \href{https://doi.org/10.1140/epjc/s10052-025-14971-8}{https://doi.org/10.1140/epjc/s10052-025-14971-8}

\bibitem{R8} Nojiri, S., Odintsov, S. D., \& Saridakis, E. N. (2019). Modified cosmology from extended entropy with varying exponent. \textit{The European Physical Journal C}, \textit{79}, 242. \href{https://doi.org/10.1140/epjc/s10052-019-6740-5}{https://doi.org/10.1140/epjc/s10052-019-6740-5}






‌



\end{thebibliography}
\end{document}